\documentclass[hyper,letterpaper,notoc]{JHEP3}

\usepackage{epsfig}
\usepackage{graphicx}

 \usepackage{amsmath}
 \usepackage{amscd}
 \usepackage{amsthm}
 \usepackage{amssymb}

\usepackage{epsf,float}

\newcommand{\be}{\begin{equation}}

\newcommand{\ee}{\end{equation}}
\newcommand{\bea}{\begin{eqnarray}}
\newcommand{\eea}{\end{eqnarray}}

\newcommand{\nn}{\nonumber}

\def\intyf{\int_{0}^{\pi R} \frac{dy}{\hat{g}_5^2}}

\def\intp{\int  \frac{d^4 p}{(2 \pi)^4}}
\def\intz{\int_{L_0}^{L_1} dz}

\def\wbrane{_{L_0}^{L_1}}

\def\cal#1{\mathcal{#1}}
\def\nott#1{\not{\!#1}}

\def\mtt#1{\mathtt{#1}}
\def\ptl{\partial}

\def\call{{\cal L}}
\def\cals{{\cal S}}

\def\calt{{\mtt t}}
\def\calb{\mtt b}

\def\Tr{\mathsf{Tr}}
\def\mtext{\mbox}

\def\eps{{\epsilon}}

\def\yyt{{\tilde{\cal Y}}}

\def\gpt{{\tilde g}^{\prime }}
\def\gptt{{\tilde g}^{\prime 2}}

\def\eps{{\epsilon}}

\def\Wt{\tilde{ {W}}}

\def\Yt{\tilde{ {Y}}}
\def\s{s_\theta}
\def\st{s_{\tilde\theta}}

\def\ct{c_{\tilde\theta}}
\def\gt{\tilde g}
\def\et{\tilde e}

\def\At{\tilde A}
\def\gammat{\tilde A}

\def\Zt{\tilde Z}

\def\piR{{\pi R}}
\def\L{{\pi R}}

\def\intyf{\int_{0}^{\pi R} \frac{dy}{\hat{g}_5^2}}

\def\intp{\int  \frac{d^4 p}{(2 \pi)^4}}
\def\intz{\int_{L_0}^{L_1} dz}

\def\wbrane{_{L_0}^{L_1}}

\title{Holographic approach to a minimal Higgsless model}

\author{R. Casalbuoni, S. De Curtis, D. Dominici\\
Department of Physics, University of Florence, and INFN, 50019 Sesto F., Firenze, Italy\\
 E-mail: \email{casalbuoni@fi.infn.it}, \email{decurtis@fi.infn.it}, \email{dominici@fi.infn.it}
}

\author{D. Dolce\\
IFAE, Universitat Aut\`onoma de Barcelona, 08193 Bellaterra, Barcelona, Spain\\
 E-mail: \email{dolce@ifae.es}}
\date{\today}
\abstract{ In this work, following an holographic approach, we carry
out a low energy effective study of a minimal Higgsless model based
on $SU(2)$ bulk symmetry broken by boundary conditions,  both in
flat and warped metric. The holographic procedure turns out to be an
useful computation technique to achieve an effective four
dimensional formulation of the model taking into account the
corrections coming from the extra dimensional sector. This technique
is used to compute both oblique and direct contributions to the
electroweak parameters in presence of fermions delocalized along the
fifth dimension. }
\begin{document}

\begin{abstract}
\noindent

\end{abstract}

\maketitle
\section{Introduction}
\label{Introduction}

 A relevant issue in the context of  high energy physics is that  extra dimensions  provide alternative ways for breaking gauge symmetries   with respect to the famous Higgs mechanism,
 \cite{Scherk:1979zr,Scherk:1978ta,Antoniadis:1990ew,Hosotani:1983xw,Hosotani:1983vn}.
  Furthermore, an additional non trivial feature of  the Yang-Mills theories in a compact extra dimension  is that the $WW$ and
  $ZZ$ elastic   scattering amplitudes can be unitarized by the tower of heavy gauge modes,
  \cite{SekharChivukula:2001hz,Chivukula:2003kq,Csaki:2003dt,Papucci:2004ip},  and the unitarity of the theory is postponed to
  a higher scale. This additional scale is related to the fact that the theory becomes strongly interacting and  the number of modes
  that can contribute to the amplitudes at high energy grows as well.
  Thus   extra dimensional models provide alternative methods with respect to the standard theory to break the gauge symmetries
  giving mass to the gauge bosons and preserve the unitarity of the $W$ and $Z$ scattering at high energy.  Since the spontaneous electroweak
  symmetry breaking via the Higgs mechanism and the unitarity restoration via the Higgs boson exchange are the main arguments
  for the existence of the Higgs boson, then the Higgs sector
  can be eliminated in favor of a compact extra dimensional sector.
The class of models, usually formulated in five dimensions  based on
  this assumption is  named Higgsless  \cite{Csaki:2003dt,Csaki:2003zu,Nomura:2003du,Barbieri:2003pr,Burdman:2003ya,Davoudiasl:2003me,Cacciapaglia:2004jz,Davoudiasl:2004pw,Barbieri:2004qk}.

Extra dimensional extensions of the Standard Model (SM) share some similarities with
strongly interacting models at effective level as can be inferred
through  the $AdS/CFT$ correspondence  \cite{Maldacena:1997re}. The
analogy between the technicolor models and  extra dimensional models
arises also by discretizing the extra dimensional theory with the
dimensional deconstruction mechanism.
In fact the deconstruction mechanism provides  a correspondence at
low energies between theories with  replicated 4D
gauge symmetries $G$   and theories with a 5D gauge
symmetry $G$ on a lattice \cite{Arkani-Hamed:2001ca,Arkani-Hamed:2001nc,Hill:2000mu,Cheng:2001vd,Randall:2002qr}. We will refer to the models with
replicated gauge symmetries as  moose models \cite{Foadi:2003xa,Hirn:2004ze,Casalbuoni:2004id,Chivukula:2004pk,Georgi:2004iy}.

 Since in the original versions of  Higgsless models the fermions coupled only with the standard gauge fields, the electroweak parameters
 $\eps_1,\eps_2, \eps_3$
or $S,T,U$, \cite{Altarelli:1993sz, Altarelli:1997et,
Peskin:1991sw},  had only oblique contributions.   These oblique
corrections tend to give large and positive contributions to the
$\epsilon_3$ (or $S$) parameter,  so that it is difficult to
conciliate the electroweak bounds with a delay of the unitarity
violation scale, \cite{Barbieri:2003pr,Georgi:2004iy}. However a
delocalization of the fermionic fields into the bulk as in
\cite{Cacciapaglia:2004rb,Foadi:2004ps}, that is allowing  standard
fermions with direct coupling to all the moose gauge fields as in
\cite{Casalbuoni:2005rs}, leads direct contributions to the
electroweak parameters that can correct the bad behavior of the
$\eps_3$ parameter.

The fine tuning which cancels out the  oblique and direct contributions to $\eps_3$ independently in each bulk point, that is from each
internal moose gauge group, corresponds to the so called ideal delocalisation of the fermions,
\cite{Casalbuoni:2005rs,SekharChivukula:2005cc,Chivukula:2005ji}.
 Therefore, it is worthwhile investigating whether it is possible to reproduce such an ideal delocalisation starting from the theoretical
 assumption of a 5D Dirac sector with an appropriate choice of the boundary conditions on the branes.

The determination of the low energy observables in  extra
dimensional models is generally  achieved  through a recursive
elimination of the heavy Kaluza Klein (KK)  excitations from the equations of
motion. However, a  much more useful way to reach the effective theory is
described in \cite{Luty:2003vm,Barbieri:2003pr,Burdman:2003ya}. This method is broadly inspired
by the holographic technique  which allows
the reduction of the
  5D theory into a four dimensional one \cite{Witten:1998qj}.

  Here we will study the
problem of the $\eps_3$ fine tuning directly in the extra
dimensional formulation of the Higgsless model using the holography
as a powerful procedure of calculus.
 In fact the bulk physics can
be taken into account by solving the bulk equations of motion with
given boundary conditions, so that one is left with a boundary or
holographic action, which is, indeed, a 4D theory
related to the original extra dimensional one.

Other solutions to get a suppressed contribution to $\eps_3$ have been investigated like
the one suggested by holographic QCD, assuming that different five dimensional metrics are
felt by the
axial and vector states \cite{Hirn:2005vk,Hirn:2006nt,Hirn:2006wg}. However recently it has been shown that the
backgrounds that allow to get a negative oblique contribution to $\eps_3$ are pathological, since require unphysical Higgs profile or higher dimensional operators  \cite{Agashe:2007mc}.

In this note we consider a 5D version of a linear
moose model previously proposed, \cite{Casalbuoni:2004id,Casalbuoni:2005rs,
 Bechi:2006sj}. The right pattern for electroweak
symmetry breaking is obtained by adding to the $SU(2)$ five
dimensional gauge symmetry, extra terms on the branes and boundary
conditions breaking the symmetry to the $U(1)_{em}$. Then we
evaluate the oblique corrections through the vacuum amplitudes of
the standard gauge bosons which can be easily obtained from the
holographic formulation. In general, the results obtained with the
holographic procedure  are in agreement with the continuum limit of
the corresponding linear moose model studied in
\cite{Casalbuoni:2005rs,Bechi:2006sj}. In presence of direct
couplings of the bulk gauge fields to standard fermions,
effective fermion current-current
interactions, which are obtained in the deconstruction analysis
\cite{Casalbuoni:2005rs}, are  recovered  from the full effective action
solving the complete
bulk equations of motion with a suitable perturbative expansion \cite{Luty:2003vm}.

In Section \ref{sectiontwo} we review the holographic description of the gauge
sector and show  how to derive the oblique contributions  to the
electroweak parameters.
In Section  \ref{sectionthree} we perform a  holographic analysis of the fermions in 5D
by solving the equations of  motion with
suitable boundary conditions and project out the bulk dynamics on
the branes, as in \cite{Contino:2004vy}.
In Section \ref{the:interaction}  we  use the bulk solutions obtained for the gauge and fermion fields in the interaction terms,
and we derive the low energy effective action from which we compute the $\eps$ parameters.
 The results obtained in the flat scenario are then extended to the warped background in Section \ref{sectionfive}.
 Conclusions are given in Section \ref{sectionsix}.

\section{Holographic analysis of the gauge sector}
\label{sectiontwo}

We review in this Section
 the continuum limit of the moose model of \cite{Foadi:2003xa, Casalbuoni:2004id,
Foadi:2004ps,Casalbuoni:2005rs,Foadi:2005hz,Bechi:2006sj} and the
holographic approach for gauge fields proposed in
\cite{Barbieri:2003pr}.

We start from an action based on a 5D Yang-Mills
theory defined on a segment, with $SU(2)$ bulk gauge symmetry and
flat metric: \be\label{flat:YM:action:gen} S^{bulk}_{YM} = -
\frac{1}{2 g_5^2}\int d^4 x \int_0^{\pi R} dy  \Tr[ F^{
MN}(x,y)F^{}_{MN}(x,y)] \,, \ee
where $g_5$ is the bulk gauge coupling
with mass dimension $-1/2$, $F^{ MN}= F^{a MN} T^a$, being $T^a$ the
generators of the $SU(2)$ symmetry such that $\Tr[T^a, T^b] =
\frac{1}{2} \delta^{a b}$, $T^a = \frac{\tau^a}{2}$ where $\tau^a$
are the Pauli matrices and  $F(A)^{a MN} =  A^{a MN} + i \epsilon^{a b
c} A_{b}^{M}A_{c}^{ N} $ with  $A^{a}_{MN} =
\partial_M A^a_N -
\partial_N A^a_M$.

We will work in the
unitary gauge $A^a_5 \equiv 0$, which  gives useful simplifications. Then integrating by parts the bulk action in
eq.~(\ref{flat:YM:action:gen}) and neglecting the trilinear and
quadrilinear couplings coming from the non-abelian terms of the
field strength, we are lead to a bilinear action written as a
boundary term plus a bulk term \cite{Barbieri:2003pr}.

From the bulk term we can get the bulk equations of motion which, in
the four momentum space, for  the transverse and longitudinal
components of the gauge field, are respectively ($\partial_5=\partial_y$):
\bea\label{eqmoto:bulk:flat} (\partial_5^2 + p^2) A_\mu^{t}(p,y) = 0
,~~~~~~~~~~~~~
\partial_5^2  A_\mu^{l}(p,y) = 0 \,.
\eea
Furthermore,  as long as one considers processes with all external
particles with mass $m_f$ much lighter than the  gauge vector mass $m_A$,
the longitudinal part of the two point function yields a suppression
of the order $(m_f/m_A)^2$. Thus, in discussing the electroweak
corrections coming from the extra gauge sector, we  will investigate only the transverse components of the
gauge field, \cite{Alam:1997nk} (for sake of simplicity from now on we will omit the
superscript).

Let us
impose the first of the eqs.~(\ref{eqmoto:bulk:flat})  as a
constraint, therefore the residual bilinear part of the 5D
 action is an holographic term, \cite{Barbieri:2003pr},
\be\label{hologr:extra:sector:SU2} \cals^{(2)holog}_{YM} = -
\frac{1}{g^2_5}  \intp Tr[A_\mu(p,y) \partial_5 A^\mu(p,y)]_0^{\pi R}\,.\ee

As said, in addition to this bilinear boundary action obtained
imposing the linear equations of motion eqs.~(\ref{eqmoto:bulk:flat}), we have the trilinear and quadrilinear
bulk terms coming from the non-abelian part of the 5D
$SU(2)$ Yang-Mills theory. Anyway they are not involved, at the leading
order,  in the electroweak parameter tree level estimation and have
weaker experimental bounds \cite{lepwwg}.

In order to solve the bulk equations of motion (\ref{eqmoto:bulk:flat})  we need to assign  boundary conditions
for each bulk field component. We will fix these conditions by requiring to recover the SM gauge content
at the extremes of the segment (branes). In order to get this, following \cite{Barbieri:2003pr},
 we
add, besides the  brane kinetic terms, mass brane terms for the gauge fields:
\bea\label{S:SM:YM:flat}
S^{brane}_{YM} &=& -  \frac{1}{2 \gt^2}\int d^4 x \int_0^{\pi R} dy \delta(y ) \Tr[ F(A)^{ \mu \nu}(x,y)F(A)^{}_{\mu \nu}(x,y)]  \nn \\
&-&  \frac{1}{4 \gptt}\int d^4 x \int_0^{\pi R} dy \delta(y - \pi R
)  F^{3 \mu \nu}(x,y) F^{3}_{\mu \nu}(x,y)\nn\\
&+&\frac{c_1^2}{2 \gt^2} \int d^4 x \int_0^{\pi R} dy \delta(y ) Tr[ (A^{ \mu} - \gt \Wt^{ \mu})( A^{ \mu} - \gt \Wt^{}_{\mu})]  \nn \\
&+&  \frac{c_2^2}{4 \gptt} \int d^4 x \int_0^{\pi R} dy \delta(y -
\pi R ) [ (A^{3 \mu} - \gpt \yyt^{ \mu})( A^{3}_{ \mu} - \gpt \yyt_{
\mu}) +  A^{1,2 \mu} A^{1,2}_{\mu}]\,, \eea
The parameters $c_{1,2}$
have the dimension of a 4D mass and in the limit
$c_{1,2} \rightarrow \infty $ fix the boundary values of the bulk
field to the standard gauge fields $\Wt_\mu=\Wt_\mu^a T^a$  and $\yyt_\mu $ (the tilde indicates unrenormalized quantities):
 \bea\label{boundcond:generic:pomarol}
 A_\mu^{\pm}(x,y)|_{y = 0} \equiv \gt {\Wt}^{\pm}_\mu(x), &&~~~ A_\mu^{\pm}(x,y)|_{y = \pi R} \equiv 0, \nn \\
 A_\mu^{3}(x,y)|_{y = 0} \equiv \gt {\Wt}^{3}_\mu(x), &&~~~ A_\mu^{3}(x,y)|_{y = \pi R} \equiv \gpt {\yyt}^{}_\mu(x)\,.
 \eea
In this way  the standard  fields are introduced as auxiliary fields. Though we are considering the flat metric case,
 these fields are  the analogous of the source fields of the $AdS/CFT$ correspondence.  Indeed, we are imposing standard gauge symmetry $SU(2)_L$
  on the $y=0$ brane, and $U(1)_Y$ on the $y=\pi R$ one.

 We are now able to write down the holographic formulation of the model by imposing the bulk equations of motion given in
 eqs.~(\ref{eqmoto:bulk:flat}) and the boundary conditions (\ref{boundcond:generic:pomarol}).
 The resulting
 Lagrangian density in momentum space at quadratic level is
 \bea\label{leff:bordo:pomarol:flat}
 \call^{(2) holog + SM }_{YM} &=& - \frac{\gpt}{2 g_5^2} \yyt^{ \mu}(p) \partial_y A^3_\mu(p,y) |_{y=\pi R} + \frac{\gt}{2 g_5^2}
 \Wt^{a \mu}(p) \partial_y A^a_\mu(p,y) |_{y= 0}  \nn \\
 &+& \frac{p^2}{2} \Wt^a_\mu(p) \Wt^{a \mu}(p) + \frac{p^2}{2} \yyt_\mu(p) \yyt^\mu(p)\,.
 \eea

Let us  comment on
 the relation between this holographic approach and the one
in which one uses the KK expansion in normal modes for the gauge
field. In this latter case the boundary conditions to be imposed are
different as can be derived by varying the bulk action with brane
kinetic terms added, \cite{Georgi:2000ks, Carena:2002me}.
 It can be shown that,
fixing the bulk field on the
boundary according to eqs.~(\ref{boundcond:generic:pomarol}) is indeed coherent with an effective description in terms
of Dirichlet and Neumann boundary conditions obtained from the
variation of an extra dimensional theory based on the bulk action
with brane kinetic terms. By expanding the bulk fields in
terms of KK eigenfunctions
\be A_\mu^a (p,y)=\sum_n f_n^a(y)
A_\mu^{a(n)}(p) \,,
\ee
after imposing Neumann conditions on both the
branes for the neutral component of the bulk field as well as
Neumann condition on the $y=0$ brane and Dirichlet condition on
$y=\pi R$ brane for the charged components of the bulk field
\cite{Foadi:2003xa}, at leading order in $\gt^2\pi R/ g_5^2$ (in
such
 a way that the  heavy non standard KK mass eigenstates can be
 neglected),
we obtain \bea\label{app:interpr:BC:barbieri}
A_\mu^{\pm}(p,0) &\sim&  f^{\pm}_0(0) W^\pm_\mu(p)  \sim \frac{\et}{\st} \Wt^\pm_\mu(p) , \nn \\
A_\mu^{\pm}(p,\pi R) &\sim&  f^{\pm}_0(\piR) W^\pm_\mu(p)  \equiv 0 , \nn \\
A_\mu^{3}(p,0) &\sim&  f^{3}_0(0) A_\mu(p) + f^{3}_1(0) Z_\mu (p) \sim \et \gammat_\mu (p)+ \et \frac{\ct}{\st} \Zt_\mu (p)=\gt \Wt^3_\mu (p)\,, \nn \\
A_\mu^{3}(p,\pi R) &\sim&  f^{3}_0(\L) A_\mu(p) + f^{3}_1(\L) Z_\mu (p)
\sim \et \gammat_\mu (p) - \et \frac{\st}{\ct} \Zt_\mu (p)=\gpt \yyt_\mu (p)\,,
\eea
where we have introduced the SM neutral fields through the standard rotation:
 $\Wt_\mu^3 = \ct \Zt_\mu  + \st \gammat_\mu \nn$,
 $\yyt_\mu = -\st \Zt_\mu  + \ct \gammat_\mu$ with $\et=\gt\st$ and $\st=\gt/\gpt$ and, again,
 the tildes are for unrenormalized quantities.

We see that the boundary values for the bulk field
expressed in
 terms of the lowest lying  KK modes in eqs.~(\ref{app:interpr:BC:barbieri})
 correspond, at effective level, to the boundary conditions given in
 eqs.~(\ref{boundcond:generic:pomarol}).
In other words, the effective holographic Lagrangian for the boundary fields $\Wt_\mu$ and
$\yyt_\mu$, which are not the mass eigenstates but linear combinations of all the KK modes,
can be used to describe the lightest states of the KK tower in the limit of heavy mass of the KK excitations.

\subsection{Precision electroweak parameters}

Let us  now start the evaluation of the oblique
corrections at tree level to the SM  precision electroweak parameters by using the holographic
Lagrangian density given in eq.~(\ref{leff:bordo:pomarol:flat}).

Writing
 the generic solutions of the bulk equations of motion in terms of the interpolating field delocalization functions $h(p,y)$, we get:
\bea\label{sol:equaz:moto}
A^{ \pm}_\mu (p,y) &=& \gt h_\pm (p, y) \Wt^{\pm}_\mu (p)\,, \nn \\
A^{ 3}_\mu (p,y) &=&  \gt h_W (p, y) \Wt_\mu^{ 3}(p) + \gpt h_Y (p,
y) \yyt_\mu(p)\nn\\&=& \et h_\gamma (p,y)\gammat_\mu(p)+\frac{\et}{\st\ct} h_Z (p,y) \Zt_\mu(p)\,,
\eea
with $h_\gamma=h_W+h_Y$ and $h_Z=\ct^2 h_W-\st^2 h_Y$.

From the boundary conditions (\ref{boundcond:generic:pomarol}) we get the boundary values for the
 functions $h(p,y)$:
\bea
&&h_\pm (p, y)\vert_{y=0}=h_W (p, y)\vert_{y=0}=1,~~~~~~~ h_Y(p, y)\vert_{y=0}=0\,, \nn\\
&&h_\pm (p, y)\vert_{y=\pi R}=h_W (p, y)\vert_{y=\pi R}=0,~~~ h_Y(p, y)\vert_{y=\pi R}=1\,.
\eea

By substituting the solutions (\ref{sol:equaz:moto}) in the holographic Lagrangian density (\ref{leff:bordo:pomarol:flat}), we can compare the result
with the generic extension of the quadratic SM gauge Lagrangian written in terms of the two point functions
\bea\label{lagra:Pi:auto} \call_{eff}^{(2)} &=& {{\gptt}} \Pi_{YY}(p^2)
\yyt_\mu \yyt^\mu + \gt^2 \Pi_{WW}(p^2) \Wt^3_\mu \Wt^{3\mu} \nn \\
&+& \gt \gpt \Pi_{W Y}(p^2) \Wt^3_\mu \yyt^{\mu} + \gt^2 \Pi_{\pm
\mp}(p^2) \Wt^\pm_\mu \Wt^{\mp \mu} \,.
\eea
Therefore we get:
\bea \label{P3YM:gener:holo} \Pi_{W Y}(p^2) = - \frac{1}{2
g_5^2}[h_Y h'_W + h_W h'_Y ]_0^{\pi R}, &&~~
\Pi_{Y Y}(p^2) = - \frac{1}{2 g_5^2}[h_Y h'_Y ]_0^{\pi R},\nn \\
\Pi_{ W W}(p^2) = - \frac{1}{2 g_5^2}[h_W h'_W  ]_0^{\pi R},&&~~~
\Pi_{\pm \mp}(p^2) = - \frac{1}{2 g_5^2}[h_\pm h'_\mp  ]_0^{\pi R}.
\eea
where $h'=\partial_y h$.

The solutions of the equations of motions (\ref{eqmoto:bulk:flat}) give
$h_\pm (p,y) \equiv h_W (p, y)$ and therefore $\Pi_{ W W}(p^2) \equiv
\Pi_{\pm \mp}(p^2)$, as a consequence of the custodial $SU(2)$
symmetry of the model.

Concerning the oblique corrections, we plug the eqs.~(\ref{P3YM:gener:holo}) in the $\epsilon$ parameter expressions given in terms of
the vacuum polarization  amplitudes
\cite{Peskin:1991sw,Altarelli:1993sz}
\bea
\epsilon_1^{oblique} &=& \gt^2\frac{  \Pi_{WW}(0)-\Pi_{\pm\mp}(0)}{{\tilde M}^2_W}\,,\\
\epsilon_2^{oblique} &=& \gt^2  \frac{d}{d p^2} (\Pi_{\pm\mp}(p^2)
-\Pi_{WW}(p^2))\vert_{p^2=0}\,,\\
\epsilon_3^{oblique} &=& \gt^2 \frac{d}{d p^2} \Pi_{WY}(p^2)\vert_{p^2=0}\,,
\eea
obtaining \bea\label{eps:hyhw:bound:flat} \epsilon_1^{oblique}
=  \epsilon_2^{oblique} = 0, ~~~~~ \epsilon_3^{oblique} =
-\frac{\gt^2}{2 g^2_5} \frac{d}{d p^2 }[h_Y h'_W + h_W h'_Y ]_{0, p^2
= 0} ^{  \pi R}\,.
\eea
This shows how $\eps_3^{oblique}$ can be
computed by knowing the wave functions of the gauge bosons and their
$y$-derivatives at the extremes of the segment.

Of course it is possible to get the two point functions
without the holographic prescription;  in this way we would
find  that the electroweak parameters can be  expressed
 as integrals along the extra dimension.
For example:

\be
\label{eps3:integr:hyhw} \epsilon_3^{oblique} =
\frac{\gt^2}{g_5^2}\int_0^{\pi R} dy   [{h}_Y {h}_W]_{p^2 = 0}\,.
\ee

Using the bulk equations of motion, the boundary conditions and
integrating by parts, this integral form for the $\eps_3^{oblique}$
parameter turns out to be  equivalent to the one expressed as
boundary terms in eq.~(\ref{eps:hyhw:bound:flat}).
 Notice that in the boundary formulation of the $\eps_3^{oblique}$  we need only the
  boundary values of the $h$ functions and their derivatives at the first order in $p^2$ whereas
  in the integral formulation we need the whole bulk profile of the  $h$ functions at zero order in $p^2$.

In the next Section we will consider the fermionic content of the model. Since fermions
are delocalized  into the bulk,
 vertex corrections occur and the direct
contributions to the electroweak parameters must be taken into
account. As a consequence, the estimation of the $\eps$ parameters is
obtained through a general formulation involving the renormalization of the
electroweak observables used in the definition of the electroweak
parameters as described in \cite{Burgess:1993vc,Anichini:1994xx}.
The new physics corrections to the quadratic part of the SM Lagrangian can be encoded in the $z$ coefficients
defined as follows:
 \bea\label{Seff:zparam:moment}
 \call_{eff}^{(2)} &=& \frac{p^2}{2}(1+ z_\gamma) \gammat_\mu \gammat^{\mu} + p^2 (1+z_W) \Wt^+_\mu
 \Wt^{- \mu} + \frac{p^2}{2}(1+ z_Z) \Zt_\mu \Zt^{\mu} - {p^2} z_{Z \gamma} \gammat_\mu \Zt^{\mu} \nn \\
 &-& \tilde{M}_W^2  \Wt^+_\mu \Wt^{- \mu} - \frac{1}{2} \tilde{M}_Z^2  \Zt_\mu \Zt^{ \mu}\,.
 \eea
 Then, comparing with  eq.~(\ref{lagra:Pi:auto}) and performing the standard change of basis in the neutral sector of the gauge fields, we can express
the $z$ corrections  in terms of the    two point functions:
 \bea\label{z:bil:vacum:prop}
 z_\gamma =  2  \et^2 \frac{d}{d p^2 }{\Pi}_{\gamma \gamma}(p^2)|_{p^2 = 0},&& ~~~
 z_W =  2 \frac{\et^2}{\st^2 } \frac{d}{d p^2 }{\Pi}_{\pm \mp}(p^2)|_{p^2 = 0}, \nn \\
 z_Z =  2 \frac{\et^2}{\ct^2 \st^2 } \frac{d}{d p^2 }\Pi_{ZZ}(p^2)|_{p^2 = 0},&& ~~~
 z_{Z \gamma} = - \frac{\et^2}{\ct \st } \frac{d}{d p^2 }\Pi_{Z \gamma}(p^2)|_{p^2 = 0},
  \eea
whereas the unrenormalized gauge boson masses
are given by
\bea\label{masses:gauge:propag}
  \tilde M^2_W  = - 2 \frac{\et^2}{\st^2 }\Pi_{\pm \mp}(0),   ~~
 \tilde M^2_Z  = - 2 \frac{\et^2}{ \ct^2 \st^2} \Pi_{ZZ}(0)\,,
  \eea
  while the photon is massless $M^2_\gamma = - 2 \et^2\Pi_{\gamma \gamma}(0) \equiv 0$.

All these parameters can be  expressed in boundary form thanks to the
holographic formulation given in eq.~(\ref{P3YM:gener:holo}).

Let us note that, since the unbroken $U_{em}(1)$ gauge symmetry guarantees that
$\Pi_{\gamma \gamma}(0)=0$, using the solutions of the bulk equations of motion and the boundary conditions for $h_W$ and $h_Y$,
 the following relation at $p^2=0$ holds:
\be\label{cond:scala:EW}
h_\gamma(0,y)= h_W(0,y) + h_Y(0,y) \equiv 1 \,.
\ee
This relation will be used in Section \ref{the:interaction} for
the derivation of the direct contributions to the precision
 electroweak parameters due to the delocalization of the fermions in
the bulk.

\subsection{Explicit calculations}

Let us now  evaluate  the
electroweak parameters with the explicit solutions of the transverse
components of the bulk gauge fields. The integral expression (\ref{eps3:integr:hyhw}) for the
 $\eps_3$ parameter makes the   analogy with the deconstruction procedure  much more direct. In fact
in the integral expression (\ref{eps3:integr:hyhw}) for the $\eps_3$ parameter, the $h$ functions are involved at the zero order in $p^2$.
So we only need to solve the second order differential equations (\ref{eqmoto:bulk:flat}) at $p^2=0$ and
to impose the boundary conditions on the branes.
The
solutions are:
  \bea\label{lead:behav:hYhW:flat:sol}
 h_Y(0,y) = \frac{y}{\pi R}, ~~~~~~~~~~
h_\pm (0,y)= h_W(0,y) = 1 - \frac{y}{\pi R} \,.
  \eea
These functions are the analogous, in the continuum limit, of the variables
$y_i$ and $z_i=1-y_i$ of the deconstructed formulation of the model given in \cite{Casalbuoni:2004id}, for which one finds
\be
\eps_3^{oblique}=\gt^2 \sum_{i=1}^K \frac{y_i}{g_i^2}(1-y_i)\,,
\label{moose}
\ee
where $K$ is the number of sites, $g_i$ are the gauge coupling constants of the replicated gauge groups
along the moose chain and $y_i=\sum_{j=1}^i f^2 /f_j^2$
with $f_j$ the link variables and $1/f^2=\sum_{j=1}^K 1 /f_j^2$.

By using in (\ref{moose}) the matching between the 5D parameters of the discretized
theory (the gauge coupling $g_5$, the lattice spacing $a$) and the
parameters of the 4D deconstructed theory (the gauge coupling
constant along the chain $g_j$, the link couplings $f_j$), namely
\cite{Bechi:2006sj}:
\be \frac{a}{g_5^2} \longleftrightarrow
\frac{1}{g_j^2}\,,~~~~~~~~~~~~~~~~~~\frac{1}{a g_5^2}
\longleftrightarrow f_j^2 \,,\ee
 and performing the continuum limit, we get the correspondence between
the two descriptions in the case of equal gauge couplings along the
chain. In such a case the variable $y_i$ is the discretized
analogous of the coordinate $y$ of the fifth dimension in the eq.~(\ref{eps3:integr:hyhw}).

By substituting eq.~(\ref{lead:behav:hYhW:flat:sol}) in (\ref{eps3:integr:hyhw}), we recover the well known result
\cite{Foadi:2003xa,Casalbuoni:2004id}:
\be\label{eps:expli:resul:flat} \epsilon_3^{oblique} =
\frac{\gt^2}{g_5^2} \frac{\pi R}{6}\,. \ee

It is  also easy to find out  the exact solutions of the bulk equations of motions for the
interpolating field delocalization functions. They take the form:
\bea\label{h:flat:pmYW} h_Y
(p, y) = \frac{\sin[p y]}{\sin[p \pi R]}, ~~~ h_\pm (p, y) = h_W (p,
y) &=&  \frac{\sin[p(\pi R - y)]}{\sin[p \pi R]}\,, \eea
from which it
is straightforward to get the same result given in eq.~(\ref{eps:expli:resul:flat}) for the electroweak parameter $\epsilon_3^{oblique}$,
by using the boundary expression
 (\ref{eps:hyhw:bound:flat}).

Moreover, by using the exact solutions  (\ref{h:flat:pmYW}) we can compute the two
point functions defined in eq.~(\ref{P3YM:gener:holo}),  and,
 by comparing with eqs.~(\ref{z:bil:vacum:prop})-(\ref{masses:gauge:propag}), we can extract
the unrenormalized  masses:
\bea\label{mass:bos:flat:tilde} \tilde{M}_Z = \frac{v}{2} \frac{\gt}{\ct}, ~~~ \tilde{M}_W =
\frac{v}{2} \gt \,, \eea
where  $ v \equiv \frac{2}{g_5
\sqrt{\pi R} } $,
and the $z$ corrections:
 \bea\label{z:par:holog:flat}
 z_\gamma =  \frac{\gt^2 \pi R \st^2}{g_5^2},&& ~~~
 z_W =  \frac{\gt^2 \pi R }{3 g_5^2},\nn \\
 z_Z = \frac{\gt^2 \pi R (\ct^4 - \ct^2 \st^2 + \s^4)}{3 \ct^2 g_5^2},&& ~~~
 z_{Z \gamma} =\frac{\gt^2 \pi R \st (-\ct^2 + \st^2)}{2 \ct g_5^2} \,.
 \eea
The above expressions  are in agreement
with those obtained by performing the continuum limit of the deconstructed moose model in
\cite{Casalbuoni:2005rs}.

Let us notice that the additional electroweak parameters introduced in \cite{Barbieri:2004qk}, namely $X,Y,W$,
 are suppressed by a  factor $M_W^2 R^2$  with respect to $S$, while $V=0$ due to the custodial symmetry of the model.

\section{ Holographic analysis of the fermionic sector}
\label{sectionthree}

For what   concerns fermions in one extra dimension we can
carry out the procedure given in  \cite{Contino:2004vy}, starting
from the following bulk action for the 5D Dirac field, in the unitary gauge ($A_5=0$):
 \be\label{action:fermion:impuls}
 S_{ferm}^{bulk}=
\frac{1}{\hat g^2_5} \int d^4x \int_0^{\pi R} dy  \left\lbrace
\bar{\Psi} \gamma^\mu D_\mu  \Psi + \frac{1}{2 }[\bar{\Psi} \gamma^5
\partial_5 \Psi - \partial_5 \bar\Psi \gamma^5 \Psi] - M \bar{\Psi}
\Psi \right\rbrace  \,,\ee
where  $\hat g_5$ is a parameter - in
general different from the  bulk gauge coupling $g_5$ - with mass
dimension $-1/2$, $M$ is a constant mass for the bulk fermions and
\be
D_\mu \Psi(x,y)= \Big(\partial_\mu +i  T_a A_\mu^a (x,y) +i Y_L A_\mu^3(x,\pi R)\Big)\Psi(x,y)\,,
\label{cova}
\ee
with  $Y_L=B-L$  the left hypercharge. The hypercharge contribution to the covariant derivative appears as a non-local
term along the extra dimension since it
is evaluated in $y=\pi R$.

 Performing the variational analysis of the
fermionic action, the bulk equations of motion for a free Dirac field
written  in terms of the left and right handed components: $\Psi = \psi_L + \psi_R$ with $\gamma_5 \psi_{L,R} = \mp
\psi_{L,R}$, are, in the momentum space:
 \bea\label{chiral:diffeq:firstord} \nott{{p}}
\psi_L(p,y) + (\partial_5 - M)\psi_R(p,y)=0 ,~~~~~~~ \nott{p}
\psi_R(p,y) - ( \partial_5 + M)\psi_L(p,y)=0 \,. \eea

The left and right
handed components of the bulk  Dirac field result  to be described
by a system of  two coupled first order differential equations that
mix the two chiral components: however
 the system can be decoupled acting on the previous eqs.~(\ref{chiral:diffeq:firstord})  with the operators $(\partial_5 +
M)$ and $(\partial_5 - M)$ respectively. Then, both right and left
handed fields satisfy the second order differential equation
\be\label{eq:moto:bulk:chir:secord} (\partial^2_5 +
\omega^2)\psi_{L,R} = 0~~  \mtext{where}~~ \omega = \sqrt{p^2 -
M^2}\,. \ee

\subsection{Boundary conditions for fermions}
\label{Boundary:condition:for:fermions}

We  generalize the procedure described in the gauge sector  to determine
 the boundary  values for bulk fermions; in fact,
following
\cite{Panico:2006em}, we add to
the bulk action (\ref{action:fermion:impuls}) the 
brane action
\bea\label{S4:Sigma:OLOGR:top} S^{brane}_{ferm} &=&
 \int d^4 x \int_{0}^{\pi R} d y  \delta(y)
 \left[ \bar{q}_L i
\gamma^\mu D_\mu q_L +  \frac{1}{   \hat g_5^2}\left( \calt_L (\bar{\psi}_R q_L +
\bar{q}_L \psi_R)- \frac{1 }{2} \bar{\Psi}\Psi \right) \right]   \nn \\
&+&  \delta(y - \pi R) \left[
\bar{q}_R i \gamma^\mu D_\mu  q_R +  \frac{1}{   \hat g_5^2}\left( \calt_R(\bar{q}_R \psi_L +
\bar{\psi}_L q_R) - \frac{1}{2} \bar{\Psi}\Psi\right)
\right] \,, \eea which contains kinetic terms for the interpolating
brane fields $q_L$ and $q_R$, their couplings to the bulk Dirac
field  $\Psi$ and pseudo-mass terms for the bulk fermion field. In agreement
with the gauge symmetries on the branes, we have
\bea
&&D_\mu q_L|_{y=0}=\left(\partial_\mu +i \gt T^a \Wt_\mu^a +i \gpt Y_L\yyt_\mu\right) q_L\,,\nn\\
&&D_\mu q_R|_{y=\pi R}=\left(\partial_\mu +i \gpt T^3 \yyt_\mu +i \gpt Y_L\yyt_\mu\right) q_R\,.
\eea

We have allowed
different couplings $\calt_{L,R}$ between the right  and left handed brane fields and the bulk fermions, \cite{Foadi:2005hz,Bechi:2006sj}.  They parameterize the delocalization in the bulk of the brane fermions and, as we will see, are responsible for the fermion masses. The couplings $\calt_{L,R}$ can in general be different for each flavor in order to reproduce the fermion mass spectrum. Since this is not
the aim of this paper, for sake of simplicity, we will assume universal $\calt_{L,R}$ and
 an implicit sum over
the flavors.

 Let us now perform the variational analysis
on the branes for the total  action $S_{ferm}^{bulk}+ S_{ferm}^{brane}$  with fixed
fields $\delta q_L = \delta q_R \equiv 0$.
The  coefficients of the variations  $\delta
\bar \psi_R$ in $y= \pi R$ and of $\delta \bar \psi_L$ in $y= 0$ are
automatically vanishing whereas the coefficients of the variation
$\delta \bar \psi_L$ in $y= \pi R$ and  $\delta \bar \psi_R$ in $y=
0$ fix the boundary values of the two chiral bulk spinors:
\be\label{BC:ferm:Lql:Rqr} \psi_L (p,0) \equiv \calt_L q_L(p), ~~~~
\psi_R (p,\pi R) \equiv \calt_R q_R(p)\,.
\ee
Thus, the degrees of
freedom in terms of which we can eliminate the bulk field in the
holographic prescription are the
 4D  fields  $q_L$ and $q_R$ which,
indeed, live on the branes $y=0$ and $y = \pi R$ respectively. Note that, with this choice,
  we get the same scenario of the moose model with
direct couplings for the fermions \cite{Casalbuoni:2005rs}, where
$q_L$ and $q_R$ correspond to the standard  left and right handed
fermions.

Once we have fixed the boundary values of the bulk fields we can  determine the explicit solutions of the differential
equations (\ref{chiral:diffeq:firstord}) with boundary conditions
(\ref{BC:ferm:Lql:Rqr}), which are
\bea\label{sol:chira:f:LR}
\psi_L(p,y) &=& f_L(p,y) \calt_L q_L(p) + {\nott p} \L\tilde{f}_L(p,y) \calt_R q_R(p), \nn \\
\psi_R(p,y) &=& f_R(p,y) \calt_R q_R(p) + {\nott{{p}}} \L
\tilde{f}_R(p,y)\calt_L q_L(p) \,,\eea
with
\bea\label{equa:mot:serone:flat} f_L(p,y) &=& \frac{ \omega
\cos[\omega (\pi R - y)] + M \sin[\omega (\pi R - y)]}
{\omega \cos[\pi R \omega ] + M \sin[\pi R \omega]}, \nn \\
\tilde f_L(p,y) &=&  \frac 1 {\pi R}\frac {\sin[\omega y]}{\omega \cos[\pi R \omega ] + M \sin[\pi R \omega]}, \nn \\
f_R(p,y) &=&\frac{  \omega \cos[\omega y] + M \sin[\omega y] }{\omega \cos[\pi R \omega] + M \sin[\pi R \omega]},\nn\\
 \tilde f_R(p,y) &=&  \frac 1 {\pi R}\frac{   \sin[\omega (\pi R - y)] }{\omega \cos[\pi R \omega] + M \sin[\pi R \omega]}\,.
 \eea
At  $O(p^2)$ we get:
\bea
 \psi_L(p,y) &\sim& \calt_L q_L(p) e^{-My}  + \nott p \calt_R q_R(p)  \frac{\sinh[M y]}{M} e^{- M \pi R}, \nn \\
\psi_R(p,y) &\sim& \calt_R q_R (p) e^{M(y-\pi R)} + \nott p \calt_L q_L(p)
\frac{\sinh[M ( y - \pi R)]}{M} e^{- M \pi R}\,. \eea
Let us
note that   we have non vanishing right and left handed
contributions in the $y=0$ and $y=\pi R$ branes respectively.
The particular solution with null bulk mass,
$M=0$, is
\bea\label{solu:ferm:mzero:psiLR} \psi_L(p,y) \sim \calt_L q_L(p) +
\nott{p} y \calt_R q_R(p), ~~~~~~~~~~ \psi_R(p,y) \sim \calt_R q_R (p)+
\nott{p} \calt_L  q_L (p)(\pi R - y). \eea Hence the two components of
the Dirac field have a flat profile along the extra dimension at
zero order in $p$. A study of the phenomenological implications of
the $M$ parameter is developed in \cite{Contino:2004vy}.

The effective Lagrangian can be deduced after the
normalization of the kinetic term as we will see in the following Section.

\section{The interaction}
\label{the:interaction}

So far we have derived the holographic description of
the fermionic sector by imposing the chiral equations of
motion (\ref{chiral:diffeq:firstord}) and the boundary conditions (\ref{BC:ferm:Lql:Rqr}), but
we have  not considered the interaction with the gauge fields. In fact, in
presence of covariant derivatives in the bulk action (\ref{action:fermion:impuls}), the interaction terms are not
eliminated by the equations of motion.
By applying the holographic  prescription, the residual fermionic
action terms of the theory are
\bea\label{interaction:exact:flat}
S_{ferm}^{holog+brane}&=& S^{brane}_{ferm} -\intp \intyf \bar{\Psi}(p,y)
\gamma^\mu\left[ {A_\mu}(p,y) + \frac{\gpt}{2} (B -L)
{\yyt}_\mu(p)\right]   \Psi(p,y)\,,\nn\\
 \eea
where   $S^{brane}_{ferm}$  is evaluated by using the (\ref{sol:chira:f:LR}) solutions.

Let us observe that the  $B-L$ interaction described in eq.~(\ref{interaction:exact:flat}), coming from the covariant derivative term (\ref{cova}),
 appears as a non local interaction term in the
fifth dimension. A way to generate this interaction through a local bulk
dynamics is to introduce an additional gauge symmetry $U(1)_{B-L}$
in the bulk with gauge coupling $g'_5$. In analogy with eq.~(\ref{S:SM:YM:flat}), the related bulk  field $B(x,y)$ is fixed on
both the boundaries by brane mass terms (in the limit of large mass)
in order to obtain the boundary conditions $ B(x,y)|_{y=0,\pi R}
=\gpt\yyt$. Since the boundary value for the $B(x,y)$ is equal on
both  branes, its delocalization   function $h_{B-L}(p,y)$ at
$p^2=0$ is flat, that is $h_{B-L}(0,y) = 1$. Therefore,  the Dirac
bulk field  has a local 5D interaction   that, at
effective level, reproduces the  one given in eq.~(\ref{interaction:exact:flat}). In fact, as we shall see in the
following analysis, the low energy effective interaction Lagrangian
at leading order in $p^2$ is described in terms of the
delocalization functions $h(0,y)$. Moreover, the holographic term of
the $U(1)_{B-L}$ bulk theory,  analogous of the one in eq.~(\ref{hologr:extra:sector:SU2}),
\be S_{B-L}^{holog}=-\frac 1 {{2
g'}_5^2}\intp
[B_\mu (p,y)\partial_5 B^\mu(p,y)]_0^{\pi R}
\ee
can be neglected if
we suppose $g'_5 \gg g_5$. In conclusion, the low energy limit of
the model with a  local bulk $U(1)_{B-L}$ interaction,  is
the same of the non local one that we have  studied. However the model
with the additional $U(1)_{B-L}$ symmetry in the bulk
and its phenomenological implications  deserve a dedicated  study, even though
the oblique corrections to the $\eps_3$ parameter are unaffected by the extra $B-L$ factor \cite{Agashe:2007mc}.

In order to evaluate the low energy effective  Lagrangian, we plug
the  generic solution of the equations of motion, eqs.~(\ref{sol:chira:f:LR}), into the eq.~(\ref{interaction:exact:flat}). Neglecting again the $p \cdot A$
terms since we are considering only the transverse components of the
bulk field, we get
 \bea\label{inter:exact:interp}
 &&{\cal L}_{ferm}^{holog+brane} =
  \frac{\pi R}{\hat g_5^2}  \calt_L^2 \tilde f_R(p,0 )
\bar q_L(p) \nott p q_L(p) +   \frac{\pi R}{\hat g_5^2}  \calt_R^2 \tilde f_L(p,\pi R)  \bar q_R(p) \nott p q_R(p) \nn \\
 &&~~~~~~~~~~~~~~~~+ \frac{\calt_L \calt_R}{2 \hat g_5^2}\left[ f_R(p,0 ) + f_L(p,\pi R )\right] (\bar q_L(p) q_R(p) + \bar q_R(p) q_L(p) )\nn\\
&&~~~~~~~~~~~~~~~~+ \bar{q}_L(p)  \gamma^\mu \left[  p_\mu - \gt  \Wt_\mu (p)-  \frac \gpt 2 (B-L) {\yyt_\mu(p)} \right]  q_L (p)\nn\\
&&~~~~~~~~~~~~~~~~+
 \bar q_R (p)\gamma^\mu  \left[  p_\mu -
\gpt  { \Yt_\mu}(p) - \frac  {\gpt(B-L)}{2} { \yyt_\mu}(p) \right]  q_R (p) \nn \\
 &&-
 \intyf  \left( \calt_L^2 f^2_L(p,y)  +
 (p \pi R)^2 \calt_L^2 {\tilde f}^2_R(p,y)\right)
 \bar   q_L(p)\gamma^\mu \left[ A_\mu (p,y) + \frac \gpt 2  (B-L) {\yyt_\mu(p)}  \right]   q_L(p)   \nn \\
 &&-\intyf   \left(\calt_R^2 f^2_R(p,y)  +  (p \pi R)^2 \calt_R^2 {\tilde f}^2_L(p,y)\right)
 \bar q_R (p)\gamma^\mu\left[  A_\mu(p,y) + \frac \gpt 2 (B-L) {\yyt_\mu(p)}  \right]   q_R(p)\,.\nn\\
\eea
 where $\Yt_\mu=\yyt_\mu T^3$.
Then we will use the generic solutions of the bulk equations of
motion for the bulk gauge  fields given  in eqs.~(\ref{sol:equaz:moto}), in
order to  deal only with the gauge fields $\Wt$ and $\yyt$.
Notice that, in presence of direct
couplings of the bulk gauge fields to standard fermions, since the
simplest holographic approach consists in solving the free equations
of motion for the fields, effective fermion current-current
interactions, which are obtained in the deconstruction analysis
\cite{Casalbuoni:2005rs}, are not recovered. The full effective action could be built solving the complete
bulk equations of motion with a suitable perturbative expansion \cite{Luty:2003vm}. As pointed out in the numerical
analysis performed in \cite{Casalbuoni:2005rs}, the current-current terms, in the region of the
parameter space allowed by the precision electroweak data, are negligible. For this reason we have only considered
the free equations of motion for the bulk fields.

Notice  also that, within  this approach, the kinetic terms for the fermions no longer come
from the bulk action but they come from the boundary kinetic terms
and from the pseudo-mass terms of the brane action $S^{brane}_{ferm}$, that
is from the terms of the type $\bar{q} \Psi$ or $\bar{\Psi} \Psi$ in
eq.~(\ref{S4:Sigma:OLOGR:top}). This implies that we have, at
$O(p^2)$, \be\label{kin:term:nonorm:int:flat} \call^{kin}_{ferm}
\sim \bar q_L  \nott p \left( 1 + \calt_L^2  \frac{\pi R}{\hat
g_5^2} \tilde f_R(0,0 )\right) q_L + \bar q_R  \nott p \left( 1 +
\calt_R^2 \frac{\pi R}{\hat g_5^2} \tilde f_L(0,\pi R )\right)  q_R
\,. \ee

In order to have canonical kinetic terms,  a normalization of the brane interpolating
fields is necessary:

 \be
\label{norm:qL:qR:ingr} q_{L} \rightarrow
\frac{q_{L}}{\sqrt{1 +  \calt_{L}^2 \frac{\pi R}{\hat g_5^2}
\tilde f_{R}(0,0 )}}\,,~~~~~~
q_R\rightarrow \frac{q_{R}}{\sqrt{1 +  \calt_{R}^2 \frac{\pi R}{\hat g_5^2}
\tilde f_{L}(0,\pi R)}}
 \,.
\ee
Using the
properties of the $f$ and $\tilde{f}$ functions given in Appendix
\ref{sec:A}, the normalization factor can be written  in
integral form
\bea
 q_{L,R} &\rightarrow&
 \frac{q_{L,R}}{\sqrt{1 + \int_0^{\pi R}dy ~b_{L,R}(y)
}} \,,
\label{norma}\eea
where
\be b_{L,R}(y) = \calt^2_{L,R}
\frac{f_{L,R}^2(0,y)}{\hat g_5^2}\,.
\label{bl}\ee

In the form (\ref{norma}) the relation between the
holographic procedure and the continuum limit of the deconstructed version of the model
 is much more evident since $\int dy\rightarrow a\sum_{i=1}^K$ ,
$b_L(y) \rightarrow b_i/a$ and $b_R(y) \rightarrow
b'_i/a$ where $a$ is the lattice spacing \cite{Casalbuoni:2005rs,Bechi:2006sj}.

After the normalization (\ref{norm:qL:qR:ingr}), defining
the electric charge as $Q = T^3 + \frac{B-L}{2}$, and using eq.~(\ref{cond:scala:EW}),
 we can extract from the effective Lagrangian (\ref{inter:exact:interp})
the lowest order interaction terms:
  \bea\label{corr:inter:bLR:flat}
 {\cal L}_{ferm} &=&
- \et Q \At_\mu\bar{q} \gamma^\mu q -    \frac{\et }{\st \ct}
{\Zt_\mu}\bar q \gamma^\mu \left\lbrace
 T^3 \frac{1- \gamma_5}{2} \left( 1 - \frac{\calb_L}{2}\right)  - T^3 \frac{1+ \gamma_5}{2}   \frac{\calb_{R}}{2} -  \st^2 Q \right\rbrace  q  \nn \\
&-& \Big [\frac{\et}{\st \sqrt 2} \Wt^-_\mu \bar{q}_d\gamma^\mu
\left\lbrace   \frac{1- \gamma_5}{2} \left( 1 -
\frac{{\calb_{L}}}{2}\right)  - \frac{1+ \gamma_5}{2}
\frac{\calb_R}{2}  \right\rbrace  q_u + h.c.\Big ]~, \eea where $q = q_L +
q_R$,
and the corrections to the electroweak currents are given by
\bea\label{def:BL:BRf} \calb_L  = \frac{ 2 \int_0^{\pi R}dy
~b_L(y) h_Y(0,y) }{1 + \int_0^{\pi R}dy ~ b_L(y) }, ~~\calb_{R}
= \frac{ 2 \int_0^{\pi R}dy
~b_R(y) h_Y(0,y) }{1 + \int_0^{\pi R}dy ~ b_R(y) }\,.
\eea

The $\calb_R$ parameter  gives rise to  charged and neutral right handed
currents  coupled with the SM gauge bosons and for this reason it
has phenomenologically strong  bounds related to the $b \rightarrow
s \gamma$ process, \cite{Larios:1999au}, and the $\mu$ decay,
\cite{Eidelman:2004fx}. Allowing different brane coupling
coefficients $\calt_L$ and $\calt_R$ for the $q_L$ and $q_R$ four
dimensional fermions, we get different values for $\calb_L$ and $\calb_R$.
In particular, a small brane coupling coefficient  $\calt_R$ with
respect to  $\calt_L$ suppress the $\calb_R$ contribution. In the following phenomenological
analysis we will neglect the $\calt_R$ contribution.

After identifying the physical parameters
as in \cite{Anichini:1994xx}
  and following the procedure used in \cite{Casalbuoni:2005rs},
that is identifying the physical fields by diagonalizing   ${\cal L}^{(2)}_{eff} $ in eq.~(\ref{Seff:zparam:moment}),
\bea \label{eq:27} \gammat_\mu =
(1-\frac{z_\gamma}{2}) A_\mu+z_{Z \gamma} Z_\mu\,,~
 \Wt^\pm_\mu = (1-\frac{z_W}{2}) W^\pm_\mu\,,~
\Zt_\mu = (1-\frac{z_Z}{2}) Z_\mu\,,~ \eea
 we can derive the expression for the  $\eps_3$ parameter, including  the oblique and direct contributions.
By taking only the leading order in $\calt_L^2 \pi R/\hat g_5^2$ and in the limit
$\gt^2\pi R/g_5^2\ll 1$, corresponding to $\gt^2/g_i^2\ll 1$ in the deconstructed
version, we get:
\be\label{eps3:flat:left} \epsilon_3 = \int_0^{\pi R} dy h_Y(0,y)
\left\lbrace  \frac{\gt^2}{g_5^2}
   h_W(0,y)  -
{b_L(y)} \right\rbrace\,.\ee
The fermion contribution  contains  $b_{L}(y)$, given in eq.~(\ref{bl}), and turns out to be proportional to the square of the left-handed fermion interpolating function at leading order in $p^2$.
Notice that, as already stated for the oblique corrections, also the direct contribution to $\eps_1$ and $\eps_2$ parameters vanishes because the corrections to the fermionic currents do not violate
the custodial $SU(2)$ symmetry of the model.

Eq.~(\ref{eps3:flat:left})
is the
continuum analogous of the $\eps_3$ parameter  found in the linear
moose with direct couplings of the left handed fermions,
\cite{Casalbuoni:2005rs}.

 The ideal fermionic delocalization, corresponding to vanishing
$\eps_3$,  that is the bulk profile $f_L(0,y)$ that makes the integrand of eq.
 (\ref{eps3:flat:left}) null in every bulk point $y$,  is related to the bulk delocalization profile $h_W$ of the $\Wt$
 interpolating field through the following relation \footnote{A similar relation  between the wave function of the ordinary fermions and the wave
 function of the standard $W$ boson is suggested  in
 \cite{SekharChivukula:2005cc}}
 \be\label{ideal:deloc:eps}
b_L(y)=\calt_L^2  \frac{f^2_L(0,y)}{\hat g_5^2} = \frac{\gt^2}{ g_5^2}
h_W(0,y) ~~~ \forall y\in[0,\L]\,.
 \ee
Using the explicit solution for $h_W$ given in eq.~(\ref{lead:behav:hYhW:flat:sol}), we find that the ideal
delocalization should be given by
\be\label{ideal:deloc:eps:ferm} f_L(0,y)
\propto \sqrt{1 - \frac y \L}\,. \ee However this ideal delocalization
is not allowed by the equations of motion for  the Dirac bulk field
independently from the assumed bulk mass as can be checked by eq. (\ref{equa:mot:serone:flat}).
This result has already been found
 in \cite{Bechi:2006sj}.

Since the ideal delocalization is not allowed, the remaining possibility to get a zero new physics contribution to $\eps_3$
is to ask for a global cancellation, that is a vanishing $\eps_3$
 without requiring  the integrand of eq.
(\ref{eps3:flat:left}) to be  zero. This links the parameters of the gauge sector to the fermionic ones
as shown in \cite{Cacciapaglia:2004rb,Foadi:2004ps,Bechi:2006sj}.

As a last point  we note that, by keeping  $\calt_R\neq 0$,  after the normalization (\ref{norma}) of the interpolating
fermionic fields, we get  the following Dirac mass term
\be\label{s:mass:tRtL} \call^{mass}_{ferm} =
\frac{1}{2}\frac{\calt_{R} \calt_L}{\hat g_5^2} \left[ \frac{f_R(0,0
) + f_L(0,\pi R )}
{ \sqrt{1 +  \int_0^{\pi R}dy~b_L(y) }
\sqrt{1 +
\int_0^{\pi R}dy~b_R(y)}} \right] (\bar q_L q_R + \bar q_R q_L
)\,. \ee
Since we are working in the limit $\calt_{L,R}^2\pi R/\hat g_5^2\ll 1$
we find that the  4D mass is  $m =
\frac{\calt_L \calt_{R}} {\hat g_5^{2}}\exp{(-M\pi R)}$. As already
noticed, \cite{Foadi:2005hz, Bechi:2006sj},
 assuming  the $\eps_3$ global cancellation, the top mass value cannot be
reproduced  without allowing for a microscopical Lorentz invariance
breaking along the extra dimension.

\section{Warped scenario}
\label{sectionfive}
\subsection{Holographic analysis for the gauge sector}

Let us now extend the holographic analysis for the gauge sector
 to the case of Randall-Sundrum (RS) metric with warp factor
$k$:
\be ds^2=\frac 1 {(k z)^2}(dx^2-dz^2)\,,
\ee
so that
\bea\label{lagr:XD:unit:warp} S_{YM}^{bulk} &=& - \frac{1}{ 2
g_5^2}\int d^4 x  \intz \left( \frac{ 1} {k z}\right) \Tr[F^{
MN}(x,z) F^{}_{MN}(x,z) ] \,, \eea
where $L_0$ and $L_1$ are the brane locations.

The bulk equations of motion in the momentum space, separating the
longitudinal and the transverse components of the gauge field,  are
\bea\label{eq:motion:warp:conf} (D_5^2 + p^2) A_\mu^{t}(p,z) = 0,
~~~~~ D_5^2  A_\mu^{l}(p,z) = 0\,,
  \eea
where
\be D_5^2 = z
\partial_5(\frac{1}{z}\partial_5)\,. \ee

Considering  only the transverse components and imposing the equations
of motion as a constraint to the 5D Yang-Mills action
in the RS metric, we are left with the holographic
Lagrangian
\be\label{leff:bordo:pomarol:warp:conf}
\call^{(2)holog}_{YM} = - \frac{1}{2 g_5^2} \frac{1}{k z} A^{a \mu}
\partial_5 A^a_\mu |_{z=L_1} + \frac{1}{2 g_5^2} \frac{1}{k z} A^{a
\mu} \partial_5 A^a_\mu |_{z= L_0}\,,
\ee
where, as in the flat case, we are neglecting the
trilinear and quadrilinear terms of this non-abelian 5D $SU(2)$ Yang-Mills theory.

For what concerns the boundary conditions we add to the warped
action in eq. (\ref{lagr:XD:unit:warp})   the same brane terms of
the flat case, given in eq. (\ref{S:SM:YM:flat}), so that in the limit $c_{1,2} \rightarrow
\infty$  we get the same boundary values (\ref{boundcond:generic:pomarol})
for the bulk field   up
to a redefinition of the brane locations
\bea\label{boundcond:generic:pomarol:warp:conf} A_\mu^{\pm}(x,z)|_{z
= L_0}\equiv \gt {\Wt}^{\pm}_\mu(x), &&~~~ A_\mu^{\pm}(x,z)|_{z
= L_1} \equiv 0, \nn \\
A_\mu^{3}(x,z)|_{z = L_0} \equiv \gt {\Wt}^{3}_\mu(x), &&~~~
A_\mu^{3}(x,z)|_{z = L_1} \equiv \gpt {\yyt}^{}_\mu(x) \,, \eea
and a new definition of the gauge couplings in order to
keep track of the metric induced on the branes: $\tilde g
\rightarrow \tilde g \sqrt{kL_0},~
 \gpt \rightarrow \gpt
\sqrt{kL_1}$.

Imposing the bulk equations of motion (\ref{eq:motion:warp:conf}), and the boundary values of the bulk
field on the branes (\ref{boundcond:generic:pomarol:warp:conf}), the holographic
formulation of the model in  warped   space-time, including also the
brane kinetic terms, is
  \bea\label{leff:bordo:pomarol:warp:conf}
\call^{(2) holog + SM }_{YM} &=& - \frac{\gpt}{2 g_5^2} \left[  \frac{1}{k z} \yyt^{ \mu}(p)
\partial_5 A^3_\mu(p,z) \right] _{z=L_1} + \frac{\gt}{2 g_5^2} \left[ \frac{1}{k z} \Wt^{a
\mu}(p) \partial_5 A^a_\mu(p,z) \right]_{z=L_0}  \nn \\
&+& \frac{p^2}{2} \Wt^a_\mu(p) \Wt^{a \mu}(p) + \frac{p^2}{2}
\yyt_\mu(p) \yyt^\mu(p)\,. \eea

By expressing the generic solutions of the bulk equations of motion in terms of the delocalization functions
as in eq. (\ref{sol:equaz:moto}),
the warped analogous of the vacuum  polarization functions, given in eqs.~(\ref{P3YM:gener:holo}),  contain the warp factor $\frac{1}{k z}$, that
is:
 \bea\label{P3YM:gener:holo:warp}
\Pi_{W Y}(p^2) = - \frac{1}{2g_5^2}\left[ \frac{1}{k z}(h_Y h'_W +
h_W h'_Y) \right] \wbrane, ~&&
\Pi_{Y Y}(p^2) = - \frac{1}{2g_5^2}\left[ \frac{1}{k z}(h_Y h'_Y) \right]\wbrane ,\nn \\
\Pi_{ W W}(p^2) = - \frac{1}{2g_5^2}\left[ \frac{1}{k z}(h_W h'_W)
\right]\wbrane,~ && \Pi_{\pm \mp}(p^2) = - \frac{1}{2g_5^2}\left[
\frac{1}{k z}(h_\pm h'_\mp)  \right]\wbrane\,, \eea
and the identity
$\Pi(p^2)_{ W W} \equiv \Pi(p^2)_{\pm \mp}$ still holds due to the custodial symmetry.
Hence, the
oblique contributions to the $\epsilon$ parameters are
\bea\label{eps:warp:h:funct} \epsilon_1^{oblique} = 0, ~~~~~
\epsilon_2^{oblique} = 0, ~~~~~ \epsilon_3^{oblique} =
-\frac{\gt^2}{2 g^2_5} \frac{d}{d p^2 }[\frac{1}{k z}(h_Y h'_W + h_W
h'_Y) ]_{L_0, p^2 = 0} ^{  L_1}\,. \eea
Also in the warped scenario the $\eps_3^{oblique}$ parameter can be
expressed in integral form, that is \be\label{eps3:integr:hyhw:warp}
\epsilon_3^{oblique} = \frac{\gt^2}{g_5^2} \int_{L_0}^{L_1} dz
\frac{1}{k z} [{h}_Y {h}_W]_{p^2 = 0}\,,\ee
for which  it is
sufficient to solve the equations  of motion at $p^2 = 0$. In this
limit the differential equation to solve is simply $ (\frac{1}{k
z}h_{Y(W)}')' \equiv 0 $, and, imposing the boundary conditions,
the solutions are
\bea\label{deloc:hyz:warp:bound} h_Y(0,z) = \frac{L_0^2 - z^2}{L_0^2
- L_1^2}, ~~~~~~~~~~~~ h_\pm(0,z) = h_W(0,z) &=&  \frac{L_1^2 -
z^2}{L_1^2 - L_0^2}\,. \eea

In order to evaluate the electroweak parameters coming from the bulk
gauge sector we need   the exact solutions
of  the equations of motion (\ref{eq:motion:warp:conf})  for the $h$ functions.
These are
given in terms of the Bessel functions $J_1$ and $Y_1$: \bea
h_Y(p,z) &=& \frac{z}{L_1} \frac{  J_1[ p z] Y_1[ p L_0]  - J_1[  p
L_0] Y_1 (p z)}{J_1 [p L_1] Y_1[ p L_0]  - J_1[p L_0]
Y_1[p L_1]} ,\nn \\
h_\pm(p,z) = h_W(p, z) &=& \frac{z}{L_0} \frac{J_1[ p z] Y_1[ p L_1]
- J_1[p L_1] Y_1[ p z]}{ J_1[ p L_0] Y_1[ p L_1]- J_1[p L_1] Y_1[p
L_0]}\,. \eea
We can now evaluate
the two point functions given in  eqs.~(\ref{P3YM:gener:holo:warp}). For instance:
\be \Pi_{WY}(p^2) = \frac{2}{k g_5^2
\pi L_0 L_1} \frac{1}{ J_1 [p L_0] Y_1[p L_1] -J_1[p L_1] Y_1[p L_0]
}\,, \ee
and from eq.~(\ref{eps:warp:h:funct}),
we get
\be
\eps_3^{oblique} =
\frac{\gt^2}{4 k g_5^2} \frac{L_1^4 - L_0^4 - 4 L_0^2 L_1^2
\log[L_1/L_0]}{(L_1^2 - L_0^2)^2}\,. \ee

Of course this result, which is in agreement with
\cite{Casalbuoni:2004id}, is the same obtained by using the zero order
solution (\ref{deloc:hyz:warp:bound}) for  the $h$ functions in
the integral form (\ref{eps3:integr:hyhw:warp}) of the $\eps_3$ parameter.

Moreover, we can evaluate the $z$  parameters, which  are needed for
the determination of the non oblique contributions to $\eps_3$
coming from the bulk Dirac sector. Plugging the vacuum amplitudes of
the warped case in eqs.~(\ref{z:bil:vacum:prop}), we obtain \bea
 z_\gamma &=&  \frac{\gt^2 \st^2}{k g_5^2} \log[\frac{L_1}{L_0}] \nn \\
 z_W &=&  -\frac{\gt^2 }{4 k g_5^2} \frac{L_0^4 - 4 L_0^2 L_1^2 + 3 L_1^4 +
              4 L_1^4 \log[L_0 / L_1]}{(L_0^2 - L_1^2)^2} ,
\nn \\
 z_Z &=& \frac{\gt^2}{4 k \ct^2 g_5^2} \frac{\left( L_0^4 - L_1^4 - 2 (L_0^2 -
L_1^2)^2 c_{2 \tilde \theta} - 4 \log[L_0/L_1]\right) (L_1^2  \ct^2 + L_0^2
\st^2)^2}{(L_0^2 - L_1^2)^2}, \nn \\
 z_{Z \gamma} &=& \frac{\gt^2 \st}{2 k \ct g_5^2} \frac{L_0^2 - L_1^2 - 2
\log[L_0/L_1](L_1^2 \ct^2 + L_0^2 \st^2)}{L_0^2 - L_1^2} \,,
 \eea
and  the unrenormalized
square masses have the same expression given in (\ref{masses:gauge:propag}) with
$ v^2 \equiv 8/(k (L_1^2 - L_0^2)
g_5^2).$

Using the leading order behavior in $p^2$ for the functions $h(p,y)$, it can
be noted that, both in  the flat and in the warped scenario,  the eqs.~(\ref{sol:equaz:moto})
reproduce the same solutions obtained with
the heavy mode elimination from the equations of motion used in the
deconstructed version of the model, \cite{Casalbuoni:2005rs},
extrapolated  to the  continuum.

\subsection{Fermions in warped scenario}

Let us now consider  fermions in the warped metric in order to find
the   holographic description  and   obtain the effective Lagrangian. Defining  $c = {M}/{k}$, the variation of the bulk
action, (for a review see for example \cite{Csaki:2005vy}),  leads to the following
bulk equations of motion for the left-handed and the right-handed  components of the Dirac field
\bea\label{eqmoto:ferm:coupled:warped} \nott{p} \psi_L(p,z) + \left(
\partial_5 - \frac{c + 2}{z}\right) \psi_R(p,z)=0,~ \nott{p}
\psi_R(p,z) - \left(  \partial_5 + \frac{c - 2}{z}\right)
\psi_L(p,z)=0 \,. \eea

As in the flat scenario, these first order differential equations can
be  decoupled in two second order differential equations, one for
the left handed spinor and one for the right handed spinor.
The solutions are given in terms of Bessel functions as
in the warped gauge sector.
The boundary conditions are fixed by adding the brane action
\bea\label{S4:Sigma:OLOGR:warp}
S^{brane}_{ferm} &=& \int d^4 x \intz
\left\lbrace \delta(z - L_0) \left[ \bar{q}_L i \gamma_\mu D^\mu q_L +
 \frac{1}{(k z)^4}\frac{1}{\hat g_5^2} \left( \calt_L(\bar{\psi}_R q_L +\bar{q}_L \psi_R) - \frac{1}{2} \bar{\Psi}\Psi\right)  \right] \right.  \nn \\
&+& \left. \delta(z - L_1)  \left[
\bar{q}_R i \gamma_\mu D^\mu q_R +  \frac{1}{(k z)^4}\frac{1}{\hat g_5^2}
\left( \calt_R(\bar{q}_R \psi_L + \bar{\psi}_L q_R) - \frac{1}{2}
\bar{\Psi}\Psi\right) \right]\right\rbrace  \,. \eea
The values of the bulk fields on the branes
are given in terms of the interpolating brane  fields
\be\label{BC:ferm:Lql:Rqr:warp} \psi_L (p,L_0) \equiv \calt_L
q_L(p), ~~~~~~~~~~~ \psi_R (p,L_1) \equiv  \calt_R q_R(p)\,. \ee

The generic solutions  for the bulk Dirac field can always be
written in the same form as for the flat  case, given in eqs.~(\ref{sol:chira:f:LR}), where the dimensional parameter $\pi R$, used
in order to give the same dimensionality to the $f$ and the $\tilde
f$, can
be thought of, for example, the characteristic length of the
extra dimension in the RS metric ($\L = L_1 - L_0$).

 In the  $p\to 0$
limit the two first order differential equations (\ref{eqmoto:ferm:coupled:warped}) are decoupled for
the left and right-handed spinors and the corresponding
delocalization functions are given by
\be\label{leadord:eqsmot:ferm}
f_L(0,z) = (\frac{z}{L_0})^{2 - c} ,~~~~~~~~~~~~~ f_R(0,z) =
(\frac{z}{L_1})^{2 + c}\,. \ee

Following the same procedure as in the flat scenario from the
interaction terms, we get:
\bea\label{interaction:exact:warped}
S_{ferm}^{holog+brane} = S^{brane}_{ferm} -   \intp \int_{L_0}^{L_1}
\frac{dz}{\hat{g}_5^2 ({k z})^4  }  \bar{\Psi}(p,z) \gamma^\mu
\left[ A_\mu (p,z)  +   \frac{\gpt}{2} (B -L) \yyt_\mu(p)\right]
\Psi(p,z)\,.\nn \\
\eea
 It is easy to prove that, at order $O(p^2)$
and with canonically normalized kinetic terms, the neutral and
charged interactions are described by the eq.
(\ref{corr:inter:bLR:flat}) where the corrections are  given by
the same expressions (\ref{def:BL:BRf}) as in  the flat case,
with

\bea
b_{L}(z)&=&\calt^2_{L}\frac {f^2_{L}(0,z)}
{\hat g_5^2{(kz)^4}}=\frac{\calt_L^2}{\hat g_5^2} \left (\frac 1{kL_0}\right )^4\left(\frac{L_0}z\right )^{2c}\,,\nn\\
b_{R}(z)&=&\calt^2_{R}\frac {f^2_{R}(0,z)}
{\hat g_5^2{(kz)^4}}=
\frac{\calt_R^2}{\hat g_5^2} \left (\frac 1{kL_1}\right )^4
\left(\frac{L_1}z\right )^{-2c}\,.
\label{blwarp}
\eea
 Neglecting the $\calt_{R}$ contribution and following the same procedure
as in the flat case,
   we find
 \be
\epsilon_3 = \intz h_Y(0,z) \left\{ \frac{1}{k z}  \frac{\gt^2}{
g_5^2}  h_W(0,z)  - b_L(z) \right\}\,. \ee
 Hence the ideal cancellation for this parameter is obtained when
\bea
\label{ideal:deloc:cond:warp}
 h_W(0,z)  = \left( \frac{g_5^2}{\gt^2}\right)  k z b_L(z)
\,.
\eea

  By considering the
behaviour in $z$ of $b_L(z)$ given in  eq. (\ref{blwarp}), and
that of $h_W(0,z)$ given in  eq.~(\ref{deloc:hyz:warp:bound}), from the
condition eq. (\ref{ideal:deloc:cond:warp}) we may argue that the
ideally delocalized  left-handed fermions could be  obtained with the
choice $c=-\frac{1}{2}$, \cite{Cacciapaglia:2004rb}. Nevertheless, because of the constant term in
$h_W$,
to satisfy exactly the condition
(\ref{ideal:deloc:cond:warp}) we must require  $L_1 = 0$ and $L_0 =  (\calt_L^2 g_5^2/(\gt^2 \hat g_5^2))^{1/3}/k$.
This  means an inversion of the branes and a singular metric on $z
=  L_1 = 0$ because of the curvature  factor  $\frac{1}{k z}$,
\cite{Bechi:2006sj}.   So, also in the case of  RS warped metric, it is not possible to link the delocalization functions of the  gauge  boson
and of the left-handed fermion in such a way to satisfy the bulk equations of motion and the ideal cancellation request.

The possibility of a global cancellation between  the gauge and fermion contributions to $\epsilon_3$ is obviously viable also for the warped metric case
\cite{Cacciapaglia:2004rb,Foadi:2004ps,Casalbuoni:2005rs,Bechi:2006sj}.

\section{Conclusions}
\label{sectionsix}

 The holographic prescription applied to the five dimensional Dirac theory, as well as to the
 five dimensional Yang-Mills theory, offers
an alternative  approach to the deconstruction analysis of the
 Higgsless models for studying low energy effective Lagrangians.
The holographic technique used here is equivalent to the elimination of the
fields of the internal sites of the moose in terms of the light
fields $\Wt$ and $\Yt$ \cite{Casalbuoni:2005rs,Bechi:2006sj}. This last
procedure can generate also current-current interactions in the low
energy Lagrangian. These terms are absent in the simplest holographic
analysis since one solves the free equations of motions for the bulk
field. However following \cite{Luty:2003vm} it could be possible  to reproduce also
the current-current interaction terms with a suitable perturbative approach.
The aforementioned  equivalence has been shown in this paper, neglecting current-current terms,  by studying a minimal Higgsless
model  based on the symmetry $SU(2)$ broken by
boundary conditions in the limit $\gt^2\pi R/g_5^2\ll 1$ which
corresponds in the deconstructed theory to the limit $\gt^2/g_i^2\ll 1$
where $g_i$ is the coupling constant of the gauge group of the $i$-th
site.

In particular we have  shown that  though an ideal delocalization
of the fermions along the extra dimension is not allowed
 by the bulk equations of motion, whatever the metric, a global cancellation of the $\eps_3$ parameter is possible, and therefore  the
 electroweak constraints  can be satisfied.
 In the 5D formulation of the model there is still an interaction
which appears to be non local in the fifth dimension. This non-locality could
be eliminated by extending the 5D symmetry to $SU(2)\times U(1)_{B-L}$
and asking for suitable boundary conditions.

\section{Acknowledgements}
D. Dolce would like to thank IFAE for hospitality and the Fondazione Angelo della Riccia for financial support.

\appendix
\section{Some useful identities}\label{sec:A}{\small

Let us write the bulk chiral fermions in terms of two degrees of freedom $q_L$ and $q_R$ with  left and right chirality respectively as
\bea\label{sol:chira:f:LR:APP}
\psi_L(p,y) &=& f_L(p,y) \calt_L q_L(p) + {\nott p} \L\tilde{f}_L(p,y) \calt_R  q_R(p), \nn \\
\psi_R(p,y) &=& f_R(p,y) \calt_R  q_R(p) + {\nott{{p}}} \L
\tilde{f}_R(p,y) \calt_L  q_L(p)\,. \eea Taking into account eqs.~(\ref{chiral:diffeq:firstord}), the functions $f_{L,R}$ and $\tilde
f_{L,R}$  satisfy the following differential equations
\begin{align}\label{sol:chira:f:LR:cond}
\calt_L  f_L + \pi R (\ptl_5 - M)  \calt_R  \tilde{f}_R &= 0, &~~~  p^2 \pi R \calt_R \tilde{f}_R - (\ptl_5 + M)\calt_L {f}_L &= 0, \nn \\
 \calt_R f_R - \pi R (\ptl_5 + M)  \calt_L  \tilde{f}_L &= 0, &~~~  p^2 \pi R \calt_L  \tilde{f}_L + (\ptl_5 - M) \calt_R f_R &= 0,
\end{align}
which  can be decoupled as \be\label{sol:chira:f:LR:cond:decoupl}
f''_{L,R} + \omega^2 f_{L,R} = 0, ~~~ \tilde{f}''_{L,R} + \omega^2
\tilde{f}_{L,R} = 0\,, \ee in analogy with eq.
(\ref{eq:moto:bulk:chir:secord}).

Furthermore multiplying the first of eqs.~(\ref{sol:chira:f:LR:cond}) by $f_L$,  integrating over $y$,  and using the third of eq.
(\ref{sol:chira:f:LR:cond})  we get the following useful identity
\be\label{corr:bulkbrane:ferm:flat:mzero} \int_0^{\pi R} d y
f_L^2(p,y) = - \pi R {[f_L(p,y) \tilde{f}_R(p,y) ]_0^{\pi R}} + (\pi
R)^2 p^2  \int_0^{\pi R} d y \tilde{f}_R^2(p,y)\,.\ee
In analogous
way by multiplying the second  of eqs.~(\ref{sol:chira:f:LR:cond})
by $f_R$, integrating over $y$,  and using the forth of eq. (\ref{sol:chira:f:LR:cond})  we
get \be\label{rel1} \int_0^{\pi R} d y f_R^2(p,y) = \pi R {[f_R(p,y)
\tilde{f}_L(p,y) ]_0^{\pi R}} + (\pi R)^2 p^2 \int_0^{\pi R} d y
\tilde{f}_L^2(p,y)\,, \ee By evaluating eq.
(\ref{corr:bulkbrane:ferm:flat:mzero})-(\ref{rel1}) at $p=0$, taking
into account the boundary conditions eqs.~(\ref{BC:ferm:Lql:Rqr})
which imply that $f_L (p,0)=f_R(p,\pi R)=1$,  we get
\bea\label{rel3}
\int_0^{\pi R} d y f_L^2(0,y) &=& - \pi R {[f_L(0,y) \tilde{f}_R(0,y) ]_0^{\pi R}} = \pi R \tilde{f}_R(0,0), \nn\\
\int_0^{\pi R} d y f_R^2(0,y) &=&  \pi R {[f_R(0,y) \tilde{f}_L(0,y)
]_0^{\pi R}} = \pi R \tilde{f}_L(0, \pi R)\,. \eea


\begin{thebibliography}{10}

\bibitem{Scherk:1979zr}
J.~Scherk and J.~H. Schwarz, ``How to get masses from extra dimensions,'' {\em
  Nucl. Phys.} {\bf B153} (1979)
61--88.

\bibitem{Scherk:1978ta}
J.~Scherk and J.~H. Schwarz, ``Spontaneous breaking of supersymmetry through
  dimensional reduction,'' {\em Phys. Lett.} {\bf B82} (1979)
60.

\bibitem{Antoniadis:1990ew}
I.~Antoniadis, ``A possible new dimension at a few TeV,'' {\em Phys. Lett.}
  {\bf B246} (1990)
377--384.

\bibitem{Hosotani:1983xw}
Y.~Hosotani, ``Dynamical mass generation by compact extra dimensions,'' {\em
  Phys. Lett.} {\bf B126} (1983)
309.

\bibitem{Hosotani:1983vn}
Y.~Hosotani, ``Dynamical gauge symmetry breaking as the Casimir effect,'' {\em
  Phys. Lett.} {\bf B129} (1983)
193.

\bibitem{SekharChivukula:2001hz}
R.~Sekhar~Chivukula, D.~A. Dicus, and H.-J. He, ``Unitarity of compactified
  five dimensional Yang-Mills theory,'' {\em Phys. Lett.} {\bf B525} (2002)
  175--182,
\href{http://www.arXiv.org/abs/hep-ph/0111016}{{\tt hep-ph/0111016}}.

\bibitem{Chivukula:2003kq}
R.~S. Chivukula, D.~A. Dicus, H.-J. He, and S.~Nandi, ``Unitarity of the higher
  dimensional Standard Model,'' {\em Phys. Lett.} {\bf B562} (2003) 109--117,
\href{http://www.arXiv.org/abs/hep-ph/0302263}{{\tt hep-ph/0302263}}.

\bibitem{Csaki:2003dt}
C.~Csaki, C.~Grojean, H.~Murayama, L.~Pilo, and J.~Terning, ``Gauge theories on
  an interval: Unitarity without a Higgs,'' {\em Phys. Rev.} {\bf D69} (2004)
  055006,
\href{http://www.arXiv.org/abs/hep-ph/0305237}{{\tt hep-ph/0305237}}.

\bibitem{Papucci:2004ip}
M.~Papucci, ``NDA and perturbativity in Higgsless models,''
\href{http://www.arXiv.org/abs/hep-ph/0408058}{{\tt hep-ph/0408058}}.

\bibitem{Csaki:2003zu}
C.~Csaki, C.~Grojean, L.~Pilo, and J.~Terning, ``Towards a realistic model of
  Higgsless electroweak symmetry breaking,'' {\em Phys. Rev. Lett.} {\bf 92}
  (2004) 101802,
\href{http://www.arXiv.org/abs/hep-ph/0308038}{{\tt hep-ph/0308038}}.

\bibitem{Nomura:2003du}
Y.~Nomura, ``Higgsless theory of electroweak symmetry breaking from warped
  space,'' {\em JHEP} {\bf 11} (2003) 050,
\href{http://www.arXiv.org/abs/hep-ph/0309189}{{\tt hep-ph/0309189}}.

\bibitem{Barbieri:2003pr}
R.~Barbieri, A.~Pomarol, and R.~Rattazzi, ``Weakly coupled Higgsless theories
  and precision electroweak tests,'' {\em Phys. Lett.} {\bf B591} (2004)
  141--149,
\href{http://www.arXiv.org/abs/hep-ph/0310285}{{\tt hep-ph/0310285}}.

\bibitem{Burdman:2003ya}
G.~Burdman and Y.~Nomura, ``Holographic theories of electroweak symmetry
  breaking without a Higgs boson,'' {\em Phys. Rev.} {\bf D69} (2004) 115013,
\href{http://www.arXiv.org/abs/hep-ph/0312247}{{\tt hep-ph/0312247}}.

\bibitem{Davoudiasl:2003me}
H.~Davoudiasl, J.~L. Hewett, B.~Lillie, and T.~G. Rizzo, ``Higgsless
  electroweak symmetry breaking in warped backgrounds: Constraints and
  signatures,'' {\em Phys. Rev.} {\bf D70} (2004) 015006,
\href{http://www.arXiv.org/abs/hep-ph/0312193}{{\tt hep-ph/0312193}}.

\bibitem{Cacciapaglia:2004jz}
G.~Cacciapaglia, C.~Csaki, C.~Grojean, and J.~Terning, ``Oblique corrections
  from Higgsless models in warped space,'' {\em Phys. Rev.} {\bf D70} (2004)
  075014,
\href{http://www.arXiv.org/abs/hep-ph/0401160}{{\tt hep-ph/0401160}}.

\bibitem{Davoudiasl:2004pw}
H.~Davoudiasl, J.~L. Hewett, B.~Lillie, and T.~G. Rizzo, ``Warped Higgsless
  models with IR-brane kinetic terms,'' {\em JHEP} {\bf 05} (2004) 015,
\href{http://www.arXiv.org/abs/hep-ph/0403300}{{\tt hep-ph/0403300}}.

\bibitem{Barbieri:2004qk}
R.~Barbieri, A.~Pomarol, R.~Rattazzi, and A.~Strumia, ``Electroweak symmetry
  breaking after LEP1 and LEP2,'' {\em Nucl. Phys.} {\bf B703} (2004) 127--146,
\href{http://www.arXiv.org/abs/hep-ph/0405040}{{\tt hep-ph/0405040}}.

\bibitem{Maldacena:1997re}
J.~M. Maldacena, ``The large $N$ limit of superconformal field theories and
  supergravity,'' {\em Adv. Theor. Math. Phys.} {\bf 2} (1998) 231--252,
\href{http://www.arXiv.org/abs/hep-th/9711200}{{\tt hep-th/9711200}}.

\bibitem{Arkani-Hamed:2001ca}
N.~Arkani-Hamed, A.~G. Cohen, and H.~Georgi, ``(De)constructing dimensions,''
  {\em Phys. Rev. Lett.} {\bf 86} (2001) 4757--4761,
\href{http://www.arXiv.org/abs/hep-th/0104005}{{\tt hep-th/0104005}}.

\bibitem{Arkani-Hamed:2001nc}
N.~Arkani-Hamed, A.~G. Cohen, and H.~Georgi, ``Electroweak symmetry breaking
  from dimensional deconstruction,'' {\em Phys. Lett.} {\bf B513} (2001)
  232--240,
\href{http://www.arXiv.org/abs/hep-ph/0105239}{{\tt hep-ph/0105239}}.

\bibitem{Hill:2000mu}
C.~T. Hill, S.~Pokorski, and J.~Wang, ``Gauge invariant effective lagrangian
  for Kaluza-Klein modes,'' {\em Phys. Rev.} {\bf D64} (2001) 105005,
\href{http://www.arXiv.org/abs/hep-th/0104035}{{\tt hep-th/0104035}}.

\bibitem{Cheng:2001vd}
H.-C. Cheng, C.~T. Hill, S.~Pokorski, and J.~Wang, ``The Standard Model in the
  latticized bulk,'' {\em Phys. Rev.} {\bf D64} (2001) 065007,
\href{http://www.arXiv.org/abs/hep-th/0104179}{{\tt hep-th/0104179}}.

\bibitem{Randall:2002qr}
L.~Randall, Y.~Shadmi, and N.~Weiner, ``Deconstruction and gauge theories in
  AdS(5),'' {\em JHEP} {\bf 01} (2003) 055,
\href{http://www.arXiv.org/abs/hep-th/0208120}{{\tt hep-th/0208120}}.

\bibitem{Foadi:2003xa}
R.~Foadi, S.~Gopalakrishna, and C.~Schmidt, ``Higgsless electroweak symmetry
  breaking from theory space,'' {\em JHEP} {\bf 03} (2004) 042,
\href{http://www.arXiv.org/abs/hep-ph/0312324}{{\tt hep-ph/0312324}}.

\bibitem{Hirn:2004ze}
J.~Hirn and J.~Stern, ``The role of spurions in Higgs-less electroweak
  effective theories,'' {\em Eur. Phys. J.} {\bf C34} (2004) 447--475,
\href{http://www.arXiv.org/abs/hep-ph/0401032}{{\tt hep-ph/0401032}}.

\bibitem{Casalbuoni:2004id}
R.~Casalbuoni, S.~De~Curtis, and D.~Dominici, ``Moose models with vanishing $S$
  parameter,'' {\em Phys. Rev.} {\bf D70} (2004) 055010,
\href{http://www.arXiv.org/abs/hep-ph/0405188}{{\tt hep-ph/0405188}}.

\bibitem{Chivukula:2004pk}
R.~S. Chivukula, E.~H. Simmons, H.-J. He, M.~Kurachi, and M.~Tanabashi, ``The
  structure of corrections to electroweak interactions in Higgsless models,''
  {\em Phys. Rev.} {\bf D70} (2004) 075008,
\href{http://www.arXiv.org/abs/hep-ph/0406077}{{\tt hep-ph/0406077}}.

\bibitem{Georgi:2004iy}
H.~Georgi, ``Fun with Higgsless theories,'' {\em Phys. Rev.} {\bf D71} (2005)
  015016,
\href{http://www.arXiv.org/abs/hep-ph/0408067}{{\tt hep-ph/0408067}}.

\bibitem{Altarelli:1993sz}
G.~Altarelli, R.~Barbieri, and F.~Caravaglios, ``Nonstandard analysis of
  electroweak precision data,'' {\em Nucl. Phys.} {\bf B405} (1993)
3--23.

\bibitem{Altarelli:1997et}
G.~Altarelli, R.~Barbieri, and F.~Caravaglios, ``Electroweak precision tests: A
  concise review,'' {\em Int. J. Mod. Phys.} {\bf A13} (1998) 1031--1058,
\href{http://www.arXiv.org/abs/hep-ph/9712368}{{\tt hep-ph/9712368}}.

\bibitem{Peskin:1991sw}
M.~E. Peskin and T.~Takeuchi, ``Estimation of oblique electroweak
  corrections,'' {\em Phys. Rev.} {\bf D46} (1992)
381--409.

\bibitem{Cacciapaglia:2004rb}
G.~Cacciapaglia, C.~Csaki, C.~Grojean, and J.~Terning, ``Curing the ills of
  Higgsless models: The $S$ parameter and unitarity,'' {\em Phys. Rev.} {\bf D71}
  (2005) 035015,
\href{http://www.arXiv.org/abs/hep-ph/0409126}{{\tt hep-ph/0409126}}.

\bibitem{Foadi:2004ps}
R.~Foadi, S.~Gopalakrishna, and C.~Schmidt, ``Effects of fermion localization
  in Higgsless theories and electroweak constraints,'' {\em Phys. Lett.} {\bf
  B606} (2005) 157--163,
\href{http://www.arXiv.org/abs/hep-ph/0409266}{{\tt hep-ph/0409266}}.

\bibitem{Casalbuoni:2005rs}
R.~Casalbuoni, S.~De~Curtis, D.~Dolce, and D.~Dominici, ``Playing with fermion
  couplings in Higgsless models,'' {\em Phys. Rev.} {\bf D71} (2005) 075015,
\href{http://www.arXiv.org/abs/hep-ph/0502209}{{\tt hep-ph/0502209}}.

\bibitem{SekharChivukula:2005cc}
R.~Sekhar~Chivukula, E.~H. Simmons, H.-J. He, M.~Kurachi, and M.~Tanabashi,
  ``Ideal fermion delocalization in five dimensional gauge theories,'' {\em
  Phys. Rev.} {\bf D72} (2005) 095013,
\href{http://www.arXiv.org/abs/hep-ph/0509110}{{\tt hep-ph/0509110}}.

\bibitem{Chivukula:2005ji}
R.~S. Chivukula, E.~H. Simmons, H.-J. He, M.~Kurachi, and M.~Tanabashi,
  ``Multi-gauge-boson vertices and chiral lagrangian parameters in Higgsless
  models with ideal fermion delocalization,'' {\em Phys. Rev.} {\bf D72} (2005)
  075012,
\href{http://www.arXiv.org/abs/hep-ph/0508147}{{\tt hep-ph/0508147}}.

\bibitem{Luty:2003vm}
M.~A. Luty, M.~Porrati, and R.~Rattazzi, ``Strong interactions and stability in
  the DGP model,'' {\em JHEP} {\bf 09} (2003) 029,
\href{http://www.arXiv.org/abs/hep-th/0303116}{{\tt hep-th/0303116}}.

\bibitem{Witten:1998qj}
E.~Witten, ``Anti-de Sitter space and holography,'' {\em Adv. Theor. Math.
  Phys.} {\bf 2} (1998) 253--291,
\href{http://www.arXiv.org/abs/hep-th/9802150}{{\tt hep-th/9802150}}.

\bibitem{Hirn:2005vk}
J.~Hirn, N.~Rius, and V.~Sanz, ``Geometric approach to condensates in
  holographic QCD,'' {\em Phys. Rev.} {\bf D73} (2006) 085005,
\href{http://www.arXiv.org/abs/hep-ph/0512240}{{\tt hep-ph/0512240}}.

\bibitem{Hirn:2006nt}
J.~Hirn and V.~Sanz, ``A negative $S$ parameter from holographic technicolor,''
  {\em Phys. Rev. Lett.} {\bf 97} (2006) 121803,
\href{http://www.arXiv.org/abs/hep-ph/0606086}{{\tt hep-ph/0606086}}.

\bibitem{Hirn:2006wg}
J.~Hirn and V.~Sanz, ``The fifth dimension as an analogue computer for strong
  interactions at the LHC,'' {\em JHEP} {\bf 03} (2007) 100,
\href{http://www.arXiv.org/abs/hep-ph/0612239}{{\tt hep-ph/0612239}}.

\bibitem{Agashe:2007mc}
K.~Agashe, C.~Csaki, C.~Grojean, and M.~Reece, ``The $S$-parameter in holographic
  technicolor models,''
\href{http://www.arXiv.org/abs/arXiv:0704.1821 [hep-ph]}{{\tt arXiv:0704.1821
  [hep-ph]}}.

\bibitem{Bechi:2006sj}
J.~Bechi, R.~Casalbuoni, S.~De~Curtis, and D.~Dominici, ``Effective fermion
  couplings in warped 5D Higgsless theories,''
\href{http://www.arXiv.org/abs/hep-ph/0607314}{{\tt hep-ph/0607314}}.

\bibitem{Contino:2004vy}
R.~Contino and A.~Pomarol, ``Holography for fermions,'' {\em JHEP} {\bf 11}
  (2004) 058,
\href{http://www.arXiv.org/abs/hep-th/0406257}{{\tt hep-th/0406257}}.

\bibitem{Foadi:2005hz}
R.~Foadi and C.~Schmidt, ``An effective Higgsless theory: Satisfying
  electroweak constraints and a heavy top quark,'' {\em Phys. Rev.} {\bf D73}
  (2006) 075011,
\href{http://www.arXiv.org/abs/hep-ph/0509071}{{\tt hep-ph/0509071}}.

\bibitem{Alam:1997nk}
S.~Alam, S.~Dawson, and R.~Szalapski, ``Low-energy constraints on new physics
  reexamined,'' {\em Phys. Rev.} {\bf D57} (1998) 1577--1590,
\href{http://www.arXiv.org/abs/hep-ph/9706542}{{\tt hep-ph/9706542}}.

\bibitem{lepwwg}
L.~O. The LEP~collaborations ALEPH, DELPHI and the LEP TGC~working group
  \href{http://www.arXiv.org/abs/LEPEWWG/TGC/2005-01}{{\tt
  LEPEWWG/TGC/2005-01}}.

\bibitem{Georgi:2000ks}
H.~Georgi, A.~K. Grant, and G.~Hailu, ``Brane couplings from bulk loops,'' {\em
  Phys. Lett.} {\bf B506} (2001) 207--214,
\href{http://www.arXiv.org/abs/hep-ph/0012379}{{\tt hep-ph/0012379}}.

\bibitem{Carena:2002me}
M.~Carena, T.~M.~P. Tait, and C.~E.~M. Wagner, ``Branes and orbifolds are
  opaque,'' {\em Acta Phys. Polon.} {\bf B33} (2002) 2355,
\href{http://www.arXiv.org/abs/hep-ph/0207056}{{\tt hep-ph/0207056}}.

\bibitem{Burgess:1993vc}
C.~P. Burgess, S.~Godfrey, H.~Konig, D.~London, and I.~Maksymyk, ``Model
  independent global constraints on new physics,'' {\em Phys. Rev.} {\bf D49}
  (1994) 6115--6147,
\href{http://www.arXiv.org/abs/hep-ph/9312291}{{\tt hep-ph/9312291}}.

\bibitem{Anichini:1994xx}
L.~Anichini, R.~Casalbuoni, and S.~De~Curtis, ``Low-energy effective lagrangian
  of the BESS model,'' {\em Phys. Lett.} {\bf B348} (1995) 521--529,
\href{http://www.arXiv.org/abs/hep-ph/9410377}{{\tt hep-ph/9410377}}.

\bibitem{Panico:2006em}
G.~Panico, M.~Serone, and A.~Wulzer, ``Electroweak symmetry breaking and
  precision tests with a fifth dimension,''
\href{http://www.arXiv.org/abs/hep-ph/0605292}{{\tt hep-ph/0605292}}.

\bibitem{Larios:1999au}
F.~Larios, M.~A. Perez, and C.~P. Yuan, ``Analysis of $tbW$ and $ttZ$ couplings
  from Cleo and LEP/SLC data,'' {\em Phys. Lett.} {\bf B457} (1999) 334--340,
\href{http://www.arXiv.org/abs/hep-ph/9903394}{{\tt hep-ph/9903394}}.

\bibitem{Eidelman:2004fx}
S.~Eidelman and J.~Hernandez, ``The $\rho(1450)$ and the $\rho(1700)$,''.

\bibitem{Csaki:2005vy}
C.~Csaki, J.~Hubisz, and P.~Meade, ``TASI lectures on electroweak symmetry
  breaking from extra dimensions,''
\href{http://www.arXiv.org/abs/hep-ph/0510275}{{\tt hep-ph/0510275}}.

\end{thebibliography}

\providecommand{\href}[2]{#2}\begingroup\raggedright\endgroup

\end{document}